\documentclass[aps,showpacs,preprint,superscriptaddress,nofootinbib]{revtex4}

\begin{document}

\title{A Canonical Analysis of the Einstein-Hilbert Action  
in First Order Form}

\author{N. Kiriushcheva}
\email{nkiriush@uwo.ca}
\affiliation{Department of Mathematics,
University of Western Ontario, London, N6A~5B7 Canada}

\author{S. V. Kuzmin} 
\email{skuzmin@uwo.ca}
\affiliation{Department of Applied Mathematics,
University of Western Ontario, London, N6A~5B7 Canada}

\author{D. G. C. McKeon}
\email{dgmckeo2@uwo.ca}
\affiliation{Department of Applied Mathematics,
University of Western Ontario, London, N6A~5B7 Canada}
\date{\today}


\begin{abstract}
Using the Dirac constraint formalism, we examine the canonical structure of the Einstein-Hilbert action $S_d = \frac{1}{16\pi G} \int d^dx \sqrt{-g} R$, treating the metric $g_{\alpha\beta}$ and the symmetric affine connection $\Gamma_{\mu\nu}^\lambda$ as independent variables. For $d > 2$ tertiary constraints naturally arise; if these are all first class, there are $d(d-3)$ independent variables in phase space, the same number that a symmetric tensor gauge field $\phi_{\mu\nu}$ possesses.  If $d = 2$, the Hamiltonian becomes a linear combination of first class constraints obeying an $SO(2,1)$ algebra. These constraints ensure that there are no independent degrees of freedom. The transformation associated with the first class constraints is not a diffeomorphism when $d = 2$; it is characterized by a symmetric matrix $\xi_{\mu\nu}$. We also show that the canonical analysis is different if $h^{\alpha\beta} = \sqrt{-g}\, g^{\alpha\beta}$ is used in place of $g^{\alpha\beta}$ as a dynamical variable when $d = 2$, as in $d$ dimensions, $\det\, h^{\alpha\beta} = - (\sqrt{-g})^{d-2}$. A comparison with the formalism used in the ADM analysis of the Einstein-Hilbert action in first order form is made by applying this approach in the two dimensional case with $h^{\alpha\beta}$ and $\Gamma_{\mu\nu}^\lambda$ taken to be independent variables.

\end{abstract}


\maketitle

\section{Introduction}
The Einstein-Hilbert action
$$S_d = \frac{1}{16\pi G}\int d^dx \sqrt{-g} R\eqno(1)$$
in $d$ dimensions is known to have an exceptionally rich and complex structure. The invariance of $S_d$ under the diffeomorphism 
$$x^{\prime\lambda} = x^{\prime\lambda}(x^\sigma)\eqno(2)$$
implies that many of the apparent degrees of freedom present in $S_d$ are in fact gauge artifacts. A Cauchy analysis of the equations of motion that follow from (1) bears this out [1]. If one were to make a weak field expansion
$$g_{\mu\nu}(x) = \eta_{\mu\nu} + G\phi_{\mu\nu}(x)\eqno(3)$$
of the metric about the Minkowski metric $\eta_{\mu\nu}$, then $S_d$ takes the form of a self interacting spin two gauge theory. The diffeomorphism invariance of eq. (2) becomes in the weak field limit
$$\phi_{\mu\nu} (x) \rightarrow \phi_{\mu\nu} (x) +\partial_\mu \theta_\nu(x) + \partial_\nu\theta_\mu(x).\eqno(4)$$
It can be shown that of the $\frac{1}{2} d(d+1)$ components of $\phi_{\mu\nu}$, there are only $\frac{1}{2} d(d-3)$ degrees of freedom that are physical, provided $d \geq 3$ [2].  These may be identified with the transverse and traceless part of $\phi_{ij}$. (Latin indices are spatial.)  If $d = 2$, there are no degrees of freedom associated with a spin two gauge field [3]. (This does not imply 
that the action is purely a surface term; if $g_{01} \neq 0$ in two dimensions 
$\sqrt{-g} R$ is no longer a total derivative \cite{KK-div}.)

Rather than analyzing $S_d$ by use of the weak field expansion of eq. (3), it is possible to employ the Dirac constraint analysis [4-6] to disentangle the canonical structure of $S_d$. This has been done in a variety of ways. Two well known papers that deal with the canonical structure of $S_4$ are [7,8]. The canonical position variables in these papers are taken to be $g_{ij}$, the metric on a spatial hypersurface. The components $g_{0\mu}$ of the metric effectively become Lagrange multiplier fields; the Hamiltonian is then a linear combination of first class constraints. The Poisson brackets of these constraints are non local and have field dependent structure constants, and hence do not form a Lie algebra [9]. The first class constraints can be used to construct the generator of a gauge transformation [10]. This generator is associated with a diffeomorphism transformation of $g_{0\mu}$ [10]; the invariance of $g_{ij}$ under a diffeomorphism transformation on a space like surface is not broken in the analysis of [7,8]. When a similar analysis is applied to $S_2$, then the algebra of 
the Poisson brackets associated with the first class constraints no longer has field dependent structure constants, but still is non local and may acquire a central charge [11]. There are other approaches to the canonical structure of $S_d$ [12-14]; these generally entail using alternate geometrical quantities as dynamical variables. We note that in these analyses, a condition has been imposed 
on the metric which may serve to eliminate a gauge invariance that would 
become apparent only if the canonical formalism is applied to the unrestricted 
action. 

It was noticed by Einstein [15] that derivations of the equations of motion in $S_d$ is simplified if $g_{\mu\nu}$ and $\Gamma_{\mu\nu}^\lambda$ are treated as being independent variables. If $d \neq 2$, the equation of motion for $\Gamma_{\mu\nu}^\lambda$ unambiguously gives 
$$\Gamma_{\mu\nu}^\lambda = \frac{1}{2} g^{\lambda\rho}\left(g_{\rho\mu , \nu} + g_{\rho \nu ,\mu} - g_{\mu\nu , \rho} \right) \equiv \left\lbrace\begin{array}{c}
\lambda \\
\mu\nu\end{array}\right\rbrace .\eqno(5)$$
(Often this approach is attributed to Palatini [16].)
If $d = 2$ then we can only say that the equation of motion for $\Gamma_{\mu\nu}^\lambda$ implies
that
$$\Gamma_{\mu\nu}^\lambda = \left\lbrace\begin{array}{c}
\lambda \\
\mu\nu\end{array}\right\rbrace + \delta_\mu^\lambda\xi_\nu +
\delta_\nu^\lambda\xi_\mu - g_{\mu\nu}\xi^\lambda\eqno(6)$$
where $\xi^\lambda$ is an undetermined vector [17,18].

Treating $g_{\alpha\beta}$ and $\Gamma_{\mu\nu}^\lambda$ as being independent can also simplify the canonical analysis of $S_d$ as then
$$\sqrt{-g} R = h^{\mu\nu} \left(\Gamma_{\mu\nu , \lambda}^\lambda - 
\Gamma_{\lambda\mu , \nu}^\lambda + 
\Gamma_{\mu\nu }^\lambda \Gamma_{\sigma\lambda}^\sigma - \Gamma_{\sigma\mu}^\lambda
\Gamma_{\lambda\nu}^\sigma \right)\eqno(7)$$
where
$$h^{\mu\nu} = \sqrt{-g}\, g^{\mu\nu}\eqno(8)$$
is the metric tensor density. (Eq. (7) can also be written in an alternate and possibly more useful form
$$\sqrt{-g} R = h^{\mu\nu} \left(G_{\mu\nu , \lambda}^\lambda + \frac{1}{d-1} G_{\lambda\mu}^\lambda G_{\sigma\nu}^\sigma - G_{\sigma\mu}^\lambda G_{\lambda\nu}^\sigma\right)\eqno(9)$$
where
$G_{\mu\nu}^\lambda = \Gamma_{\mu\nu}^\lambda - \frac{1}{2} \left(\delta_\mu^\lambda \Gamma_{\sigma\nu}^\sigma + \delta_\nu^\lambda \Gamma_{\sigma\mu}^\sigma\right)$.) It is apparent from eq. (7) that the Einstein-Hilbert Lagrangian is at most linear in all derivatives with this choice of independent variables.

In the next section we examine the canonical structure of eq. (7) using the Dirac constraint formalism [4-6].  For $d \geq 3$, some of the components of $\Gamma_{\mu\nu}^\lambda$ which are independent of time derivatives enter $L$ nonlinearly so it is hard to completely disentangle the constraint structure.

However, if $d = 2$ the Hamiltonian associated with $L$ in (7) reduces to a linear combination of constraints.  The actual constraint structure when $d = 2$ is contingent upon whether $g^{\alpha\beta}$ or $h^{\alpha\beta}$ is treated as independent, as by eq. (8)
$$\det\, h^{\alpha\beta} = -\left(\sqrt{-g}\right)^{d-2} .\eqno(10)$$
showing that when $d = 2$ it is not possible to express $g^{\alpha\beta}$ in terms of $h^{\alpha\beta}$.
In both cases though the net number of degrees of freedom is zero. The quantization of this topological action is treated in [19,20]. We discuss the canonical structure of $S_2$ in first order form in section three, showing how all degrees of freedom are constrained. (In [21,22], gravity in two dimensions is treated as being over constrained.) 

The canonical structure of the Einstein-Hilbert action described by [7,8] is derived from the first order form of this action in $d = 4$ dimensions in [23,24]. In order to demonstrate how this ADM analysis differs from what is being proposed here, we apply the ADM analysis to $S_2$ in section four, contrasting it to what is outlined in section three.

\section{The Einstein-Hilbert Lagrangian in First Order Form}

Though it is not necessary to do so, it is convenient to transfer derivatives occurring in $\sqrt{-g} R$ as given in eq. (7) from the affine connections so that
$$S_d = \int d^dx \left[ \left(-\Gamma_{ij}^0 \dot{h}^{ij} \right) + \left(\Gamma_{ji}^i - \Gamma_{j0}^0\right) \dot{h}^{j} + \Gamma_{0i}^i \dot{h} - \Phi \left(h^{\alpha\beta}, \Gamma_{\mu\nu}^\lambda\right)\right] .\eqno(11)$$
(Latin indices are spatial, $\dot{f}$ denotes $\partial_tf$ and $h = h^{00}, h^i = h^{0i}$.) Denoting momenta conjugate to $h$ by $\pi$ and to $\Gamma$ by $\Pi$, we have the primary constraints
$$\Pi_\lambda^{\mu\nu} = 0,\;\;\pi_{ij} = -\Gamma_{ij}^0, \;\; \pi_j = \Gamma_{ji}^i - \Gamma_{j0}^0, \;\;\pi = \Gamma_{0i}^i\;.\eqno(12-15)$$
If now
$$\Lambda = \Gamma^0_{00},\;\;\;\;\;\Lambda^k = \Gamma_{00}^k, \;\;\;\;\; \Sigma_i = \Gamma_{ij}^j \;\;. \eqno(16-18)$$
$$\Sigma_j^i = \Gamma_{j0}^i - \frac{1}{d-1} \left(\Gamma_{k0}^k \delta_j^i\right),\;\;\;\;\; \Sigma_{jk}^i = \Gamma_{jk}^i - \frac{1}{d} \left(\delta_j^i \Gamma_{k\ell}^\ell + \delta_k^i \Gamma_{j\ell}^\ell\right),\eqno(19-20)$$
then the function $\Phi$ in eq. (11) becomes
$$\Phi = \Lambda^k\left(h_{,k} - h\pi_k - 2h^i\pi_{ik}\right)
+\Lambda \left(-h_{,i}^i - \pi h+ h^{ij}\pi_{ij}\right)
+ h\left(\Sigma_j^i \Sigma_i^j + \frac{1}{d-1} \pi^2\right)\nonumber$$
$$+ h^i\left(-2\Sigma_{i,k}^k + \frac{d-3}{d-1} \pi_{,i} + 2\pi\pi_i - 2\pi\Sigma_i + 2\Sigma_\ell^k\Sigma_{ki}^\ell
 - 2 \frac{(d-1)}{d} \Sigma_k\Sigma_i^k\right)\nonumber$$
$$+ h^{ij}\left(-\Sigma_{ij,k}^k + \frac{2(d-1)}{d} \Sigma_{i,j}- \pi_{i,j} + \pi_i\pi_j
- 2 \Sigma_i^k\pi_{kj} + \Sigma_{\ell i}^k\Sigma_{kj}^\ell + \pi_k \Sigma_{ij}^k\right. \nonumber$$
$$\left. -2\frac{(d-1)}{d} \Sigma_i\pi_j + \frac{(d-1)(d-2)}{d^2} \Sigma_i\Sigma_j + \left(\frac{d-3}{d-1}\right)\pi\pi_{ij} - 2 \frac{(d-1)}{d} \Sigma_k\Sigma_{ij}^k\right).\eqno(21)$$
The fields $\Lambda$, $\Lambda^k$, $\Sigma_i$, $\Sigma_j^i$ and $\Sigma_{jk}^i$ all have vanishing conjugate momenta which constitute a set of additional primary constraints. Associated with these primary constraints are a set of secondary constraints. The secondary constraints associated with the momenta conjugate to the fields $\Lambda^k$ and $\Lambda$ are
$$\chi = h_{,i}^i + h\pi - h^{ij}\pi_{ij},\eqno(22)$$
$$\chi_k = h_{,k} - h\pi_k - 2h^i\pi_{ik}.\eqno(23)$$
The usual Poisson brackets (PB) imply that these satisfy the closed algebra
$$\left\lbrace \chi_i,\chi\right\rbrace = \chi_i\;\;\;\;\;\;
\left\lbrace \chi_i,\chi_j\right\rbrace = 0 .\eqno(24,25)$$

The fields $\Sigma_i$, $\Sigma_j^i$ and $\Sigma_{jk}^i$ (collectively denoted by $\vec{\Sigma}$) have equations of motion that lead to additional constraints when $d > 2$. These fields enter into $\Phi$ in such a way that $\Phi$ can be schematically written as
$$\Phi = \vec{\Lambda} \cdot \vec{\chi} + \frac{1}{2} \vec{\Sigma}^T \tilde M
\vec{\Sigma} + \vec{\Sigma} \cdot \vec{v} + Z.\eqno(26)$$
The momenta conjugate to the fields which contribute to $\vec{\Sigma}$ vanish; these primary constraints lead to secondary constraints of the form
$$
\tilde M \vec{\Sigma} + \vec{v} = 0.\eqno(27)$$
From eq. (27), it is evident that if $\vec{\Sigma}$ has $s$ components and if the rank of the matrix $\tilde M$
is $r$, then eq. (27) constitutes a set of $r$ second class constraints and $s - r$ first class constraints. To determine the rank of $\tilde M$,
we consider that portion of $\Phi$ given in eq. (21) that is bilinear in $\vec{\Sigma}_1$:
$$\Phi_2 = h \Sigma_j^i\Sigma_i^j + 2 h^i\left(\Sigma_\ell^k \Sigma_{ki}^\ell - \frac{(d-1)}{d}\Sigma_k\Sigma_i^k\right)$$ 
$$+ h^{ij}\left(\Sigma_{\ell i}^k \Sigma_{kj}^\ell + \frac{(d-1)(d-2)}{d^2} \Sigma_i\Sigma_j- \frac{2(d-1)}{d} \Sigma_k \Sigma_{ij}^k\right).\eqno(28)$$
In order to diagonalize $\Phi_2$ in the fields $\Sigma_i$, $\Sigma_j^i$ and $\Sigma_{jk}^i$, we make the transformations
$$\Sigma_i \rightarrow A h \Sigma_i\eqno(29)$$
$$\Sigma_j^i \rightarrow B h \Sigma_j^i + C\left(h^i \Sigma_j - \frac{1}{d-1} \delta_j^i h^k \Sigma_k\right) + D h^k \Sigma_{jk}^i\eqno(30)$$
$$\Sigma_{jk}^i \rightarrow E h \Sigma_{jk}^i .\eqno(31)$$
These transformations are consistent with $\Sigma_j^i$ and $\Sigma_{jk}^i$ being traceless. With the choice $D = -E$, $C = \frac{d-1}{d} A$ and setting $q^{ij} = h^2 h^{ij} - h h^i h^j$, we find that eq. (28) reduces to
$$\Phi_2 = B^2 h^3 \Sigma_j^i \Sigma_i^j + q^{ij}\left[\frac{d-2}{d-1} C^2 \Sigma_i\Sigma_j + D^2 \Sigma_{\ell i}^{k} \Sigma_{kj}^\ell + 2C D \Sigma_k\Sigma_{ij}^k\right],\eqno(32)$$
which, with $\tilde{\Sigma}_{jk}^i =D\Sigma_{jk}^i + C\left(\delta_j^i \Sigma_k + \delta_k^i\Sigma_j\right)$,
becomes
$$= B^2 h^3 \Sigma_j^i\Sigma_i^j + q^{ij}\left(\tilde{\Sigma}_{\ell i}^k
\tilde{\Sigma}_{kj}^\ell - \frac{d^2}{d-1} C^2 \Sigma_i\Sigma_j\right).\eqno(33)$$
It is evident once $\Phi_2$ is written in terms of these variables that the matrix $\tilde M$
of eq. (26) is invertible (i.e., the rank of this matrix equals its dimension). Consequently, all of the equations of motion associated with the variables $\Sigma_i$, $\Sigma_j^i$ and $\Sigma_{jk}^i$ are second class constraints.

We now consider the number of variables and number of constraints we have encountered. There are $\frac{d(d+1)}{2}$ components of $h^{\mu\nu}$ and $\frac{d^2(d+1)}{2}$
components of $\Gamma_{\mu\nu}^\lambda$; in total then there are 
$\frac{d(d+1)^2}{2}$ independent fields. By eqs. (13-15), the momenta $\pi$, $\pi^i$ and $\pi^{ij}$ and the momenta $\Pi_0^{ij}$, $\left(\Pi_i^{ji} - \Pi_0^{j0}\right)$, $\Pi_i^{i0}$ together generate $d(d+1)$ second class constraints, while the vanishing of momenta $\Pi_0^{00}$, $\Pi_k^{00}$ are by eqs. (16,17) a set of $d$ first class constraints. In total, there are
$[d-1] + [d(d-2)] + [\frac{1}{2} (d-2)(d^2-1)] = \frac{1}{2} d(d^2 - 3)$ variables $\Sigma_i$, $\Sigma_j^i$, $\Sigma_{jk}^i$; together the equations of motion of the variables $\vec{\Sigma}$ and momenta associated with these $\vec{\Sigma}$ make up $d(d^2 - 3)$ second class constraints.

We now assume that the $d$ constraints $\chi$, $\chi_k$ of eqs. (22,23) are all first class and that these secondary constraints are responsible for a further set of $d$ tertiary constraints which are themselves first class, and which do not in turn generate further constraints. (Having $\chi$, $\chi_k$ being first class is consistent with eqs. (24,25) but it is necessary to see if they have vanishing PB with the remaining secondary constraints.) In total then there are $d + d + d = 3d$ first class constraints and $d(d+1) + d(d^2-3) = d^3 + d^2 - 2d$ second class constraints. As these $3d$ first class constraints require the introduction of $3d$ additional constraints (the ``gauge conditions''), there would now be in total $3d + 3d + (d^3 + d^2 - 2d) = d(d^2 + d + 4)$ restrictions on the $d(d+1)^2$ variables in phase space (the $h$'s, $\Gamma$'s and their associated momenta). This would mean that there are $d(d+1)^2 - d(d^2 + d + 4) = d(d-3)$ independent variables in phase space. As has been noted in the introduction, this is the number of degrees of freedom associated with a spin two gauge field in $d$ dimensions. Consequently, although we have not explicitly proven our assumptions about the secondary and tertiary constraints, we have demonstrated that our assumptions are consistent in that the number of physical degrees of freedom in $d$ dimensions is the same as are present in a $d$ dimensional spin two gauge theory.

It is possible that the sequence of $3d$ first class constraints ($d$ primary, $d$ secondary and $d$ tertiary) is associated with the $d$ parameters that characterize the full four dimensional diffeomorphism invariance of (1); the generator of this transformation can be found using the formalism of [10].  The generator would involve $d$ parameters on account of there being $d$ primary first class constraints, as well as the first and second time derivatives as there are $d$ secondary and $d$ tertiary first class constraints following from these primary constraints.  The diffeomorphism transformation of eq. (2) gives rise to second derivatives of the gauge parameters when the transformations of $g^{\alpha\beta}$ and $\Gamma^\lambda_{\mu\nu}$ are considered.  However, we are open to the possibility that this procedure in fact leads to an invariance of the action of eq. (1) that does not correspond to a diffeomorphism, as this is in fact what happens in $d = 2$ dimensions, as will be demonstrated below. In Ref. 25 it was 
noted that the portion of Eq. (7) that is bilinear in $g$ and $\Gamma$ 
possesses an unusual invariance.

We will now turn our attention to $d = 2$ dimensions where the constraint structure inherent in the Einstein-Hilbert action is more easily disentangled as the Hamiltonian is now a linear combination of first class constraints.

\section{The Two Dimensional Action}

As has been emphasised in a number of papers [17,18,25,26] the first order formalism when applied to the two dimensional Einstein-Hilbert action has some peculiar features. In dimensions higher than two, either the metric $g_{\alpha\beta}$ or the metric density $h^{\alpha\beta}$ can be used as dynamical variables as eq. (8) allows to pass from one to the other.  However, in two dimensions, if $h^{\alpha\beta}$ is used as a dynamical variable, then by eq. (10) we must have $\det\, h^{\alpha\beta} = -1$ in order for the theory to be a metric theory. This can be ensured by supplementing the action of eq. (7) with a Lagrange multiplier term
$$\lambda \Xi \equiv -\lambda(\det\, h^{\alpha\beta} + \rho).\eqno(34)$$
The factor of $\rho$ rather than 1 in eq. (34), as would be implied by eq. (10), is a reflection of the fact that the Einstein-Hilbert action is invariant under a rescaling of $h^{\alpha\beta}$.  Eventually, the condition $\rho = 1$ can be imposed.

We now will examine how the Dirac constraint formalism can be applied to $S_2$, both when using $h^{\alpha\beta}$ and $\Gamma_{\mu\nu}^\lambda$ as independent fields and when using $g^{\alpha\beta}$ and $\Gamma_{\mu\nu}^\lambda$.
The former approach has been considered in ref. [27].

The great simplification that occurs in two dimensions is that the traceless quantities $\Sigma_j^i$ 
and $\Sigma_{jk}^i$ of eqs. (18-20) vanish. Hence, for $d = 2$, we find that eq. (11) reduces to simply
$$S_2 =\int d^2x \left[\pi_{11}\dot{h}^{11} + \pi_1\dot{h}^1 + \pi \dot{h}\right.\eqno(35)$$
$$\left. - \zeta_1\chi^1 - \zeta^1\chi_1 - \zeta\chi\right]\nonumber$$
where
$$\pi_{11} = -\Gamma_{11}^0\, ,\;\;\;\;\; \pi_1 = \Gamma_{11}^1 - \Gamma^0_{10}\, ,\;\;\;\;\;\pi = \Gamma^1_{10},\eqno(36-38)$$
$$\zeta^{1} = \Gamma_{00}^1\, ,\;\;\;\;\; \zeta_1 = -\Gamma_{10}^0\, ,  \;\;\;\;\;\zeta = -\Gamma^0_{00} + \Gamma_{01}^1\eqno(39-41)$$
and
$$\chi_1 = h_{,1} - h \pi_1 - 2h^1\pi_{11},\eqno(42)$$
$$\chi^1 = h^{11}_{,1} + 2h^1 \pi + h^{11}\pi_1,\eqno(43)$$
$$\chi = h^1_{,1} + h \pi - h^{11}\pi_{11}.\eqno(44)$$
From eq. (35), we see that the momenta conjugate to the $h's$ are given by the $\pi$'s of eqs. (36-38); we thus have a set of six primary second class constraints. There are in addition three primary first class constraints associated with the momenta $\Pi_1$, $\Pi^1$ and $\Pi$ conjugate the variables $\zeta^1$, $\zeta_1$ and $\zeta$ respectively of eqs. (39-41). These in turn lead to the secondary constraints $\chi^1$, $\chi_1$ and $\chi$, all of which are first class as their PB satisfy the algebra
$$\left\lbrace \chi , \chi_1 \right\rbrace = -\chi_1 , \;\;\;\;\;\left\lbrace \chi , \chi^1 \right\rbrace = \chi^1 ,\;\;\;\;\; \left\lbrace \chi_1 , \chi^1 \right\rbrace = 2\chi .\eqno(45-47)$$
If $\sigma_a = \frac{1}{2} \left(\chi^1 + \chi_1\right)$, $\sigma_b = \frac{1}{2} \left(\chi^1 - \chi_1\right)$, $\sigma_c = \chi$, then
$$\left\lbrace \sigma_a , \sigma_b \right\rbrace = \sigma_c ,\;\;\;\;\;
\left\lbrace \sigma_c , \sigma_a \right\rbrace = \sigma_b ,\;\;\;\;\;\left\lbrace \sigma_b , \sigma_c \right\rbrace = -\sigma_a \eqno(48-50)$$
which, upon replacing the classical PB by $(-i)$(quantum commutator) becomes the Lie algebra of $SO(2,1)$.

In order to ensure that our theory is a metric theory, we must supplement our Hamiltonian with eq. (34) so that now
$$H = \zeta_1 \chi^1 + \zeta^1 \chi_1 + \zeta\chi + \lambda(hh^{11} - (h^1)^2 + \rho).\eqno(51)$$
If $p_\lambda$ and $p_\rho$ are momenta conjugate to $\lambda$ and $p$ respectively, we find that there are now two additional constraints
$$p_\lambda = 0 = p_\rho\eqno(52,53)$$
as well as the secondary constraint
$$\Xi \equiv h h^{11} - (h^1)^2 + \rho = 0.
\eqno(54)$$
We can show that
$$\left\lbrace \chi^1, \Xi\right\rbrace = \left\lbrace \chi_1 , \Xi\right\rbrace = \left\lbrace \chi , \Xi \right\rbrace = 0.\eqno(55)$$
It is apparent that $p_\lambda = 0$ is a first class constraint while $p_\rho = \Xi = 0$ are a pair of second class constraints. We note that there are in total eleven fields in the model ($h^{\alpha\beta}, \Gamma_{\mu\nu}^\lambda , \lambda , \rho$), seven first class constraints, which in turn imply seven gauge conditions, as well as  eight second class constraints. As a result, all 22 degrees of freedom present in phase space are constrained.  This is consistent with there being no physical degrees of freedom in the model. Treatments of this Lagrangian appear in [19,20].

The approach of ref. [10] can now be used to find the generator of the gauge transformation associated with the first class constraints in the system described by eq. (35). This involves classifying the first class constraints as being either primary or secondary; the primary first class constraints are
$$C_p^a = (\Pi_1, \Pi^1, \Pi)\eqno(56)$$
with the corresponding secondary first class constraints being
$$C_s^a = (-\chi^1, -\chi_1, -\chi).\eqno(57)$$
Finding a generator $G$ of a transformation that leaves the action invariant involves first setting
$$G_{(1)}^a (x) = C_p^a (x)\eqno(58)$$
and then examining
$$G_{(0)}^a(x) = -C_s^a(x) + \int dy\,\alpha^a_{\;\;b} (x,y) C_p^b(y).\eqno(59)$$
The functions of $\alpha^\alpha_{\;\;b}(x,y)$ are determined by the condition $\left\lbrace G_{(0)}^a, H\right\rbrace = 0$. The generator is then given by 
$$G = \epsilon_a G_{(0)}^a + \dot{\epsilon}_a G_{(1)}^a .\eqno(60)$$
Together, eqs. (51), (56) and (57) lead to
$$G_{(1)}^a = \left(\chi^1 - \zeta\Pi^1 -2\zeta^1\Pi ,\;\;\; \chi_1 + 2\zeta_1\Pi + \zeta\Pi_1 ,\;\;\; \chi - \zeta^1\Pi_1 + \zeta_1\Pi^1\right)\eqno(61)$$
so that the generator of eq. (60) becomes
$$G = \int\left[\epsilon_1\left(\chi^1 - \zeta\Pi^1 -2\zeta^1\Pi\right) + \epsilon^1\left( \chi_1 + 2\zeta_1\Pi + \zeta\Pi_1\right)\right. \eqno(62)$$
$$\left. + \epsilon\left( \chi - \zeta^1\Pi_1 + \zeta_1\Pi^1\right) + \dot{\epsilon}_1 
\Pi^1 + \dot{\epsilon}^1\Pi_1 + \dot{\epsilon}\Pi\right]dx.\nonumber$$
The change generated by $G$ in a quantity $A$ is given by $\left\{A, G\right\}$. From this, we find that eq. (62) leads to by eqs. (42-44),
$$\delta h = 2\epsilon_1 h^1 + \epsilon h,\eqno(63)$$
$$\delta h^1 = -\epsilon^1 h + \epsilon_1 h^{11},\eqno(64)$$
$$\delta h^{11} = -2\epsilon^1 h^1 - \epsilon h^{11},\eqno(65)$$
$$\delta \pi = \epsilon_{,1}^1+  \epsilon^1 \pi_1 - \epsilon\pi ,\eqno(66)$$
$$\delta \pi_1 = \epsilon_{,1} -  2\epsilon_1 \pi + 2\epsilon^1\pi_{11},\eqno(67)$$
$$\delta \pi_{11} = \epsilon_{1,1} -  \epsilon_1 \pi_1 + \epsilon\pi_{11},\eqno(68)$$
$$\delta \zeta^1 = \epsilon^1\zeta -  \epsilon \zeta^1 + \dot{\epsilon}^1,\eqno(69)$$
$$\delta \zeta_1 = -\epsilon_1\zeta +  \epsilon \zeta_1 + \dot{\epsilon}_1,\eqno(70)$$
$$\delta \zeta = -2\epsilon_1\zeta^1 +  2\epsilon^1 \zeta_1 + \dot{\epsilon}.\eqno(71)$$
From eqs. (36-41), eqs. (66-71) lead to 
$$\delta\Gamma_{00}^1 = \dot{\epsilon}^1 + \epsilon^1 \left(\Gamma_{01}^1 - \Gamma_{00}^0\right) - \epsilon\Gamma_{00}^1,\eqno(72)$$
$$\delta\Gamma_{01}^0 = -\dot{\epsilon}_1 + \epsilon_1 \left(\Gamma_{01}^1 - \Gamma_{00}^0\right) + \epsilon\Gamma_{01}^0,\eqno(73)$$
$$\delta\Gamma_{11}^0 = -\epsilon_{1,1} + \epsilon_1 \left(\Gamma_{11}^1 - \Gamma_{01}^0\right) + \epsilon\Gamma_{11}^0,\eqno(74)$$
$$\delta\Gamma_{01}^1 = \epsilon_{,1}^1 + \epsilon^1 \left(\Gamma_{11}^1 - \Gamma_{01}^0\right) - \epsilon\Gamma_{01}^1,\eqno(75)$$
$$\delta\Gamma_{11}^1 = -\dot{\epsilon}_1 + \epsilon_{,1} -2\epsilon^1\Gamma_{11}^0 + \epsilon_1\left(-\Gamma_{01}^1 - \Gamma_{00}^0\right) + \epsilon\Gamma_{01}^0,\eqno(76)$$
$$\delta\Gamma_{00}^0 = -\dot{\epsilon} + \epsilon_{,1}^1 + \epsilon^1\left(\Gamma_{11}^1 + \Gamma_{01}^0\right) + 2\epsilon_1 \Gamma_{00}^1 -  \epsilon\Gamma_{01}^1.\eqno(77)$$
It is apparent just from the fact that the transformations of eqs. (63-65), (72-77) are characterized by three parameters ($\epsilon^1, \epsilon_1, \epsilon$) that these transformations do not correspond to a diffeomorphism in two dimensions, as this transformation is associated with only two parameters.  Indeed, if we set
$$\xi_{00} = -\epsilon^1,\;\;\;\;\; \xi_{11} = \epsilon_1,\;\;\;\;\; \xi_{01} = \xi_{10} = -\frac{1}{2}\epsilon,\eqno(78-80)$$
then eqs. (63-65), (72-77) can be expressed as
$$\delta h^{\alpha\beta} = -\left(\epsilon^{\alpha\lambda} h^{\sigma\beta} + \epsilon^{\beta\lambda} h^{\sigma\alpha}\right)\xi_{\lambda\sigma},\eqno(81)$$
$$\delta\left[\Gamma_{\mu\nu}^\lambda - \frac{1}{2} \left(\delta_\mu^\lambda \Gamma_{\sigma\nu}^\sigma + \delta_\nu^\lambda \Gamma_{\sigma\mu}^\sigma
\right)\right]\nonumber$$
$$= \epsilon^{\lambda\sigma} D_\sigma (\Gamma)\xi_{\mu\nu} + \epsilon^{\lambda\sigma}\Gamma_{\rho\sigma}^\rho \xi_{\mu\nu}\eqno(82)$$
where $\epsilon^{01} = -\epsilon^{10} = 1$, $D_\rho\xi_{\mu\nu} = \partial_\rho\xi_{\mu\nu} - \Gamma_{\rho\mu}^\sigma \xi_{\sigma\nu} - \Gamma^\sigma_{\rho\nu}\xi_{\mu\sigma}$.
These transformations are quite similar to those noted in [25] for two dimensional gravity in the first order formalism.

From eq. (62) we also see that eq. (10) is respected as 
$$\left\lbrace \Xi , G \right\rbrace = 0.
\eqno(83)$$
Furthermore if $L$ is taken to be given by the integrand of eq. (35), then
$$\left\lbrace L, G \right\rbrace = \dot{h}^1 \epsilon_{,1} - h^1_{,1} \dot{\epsilon} + \dot{h}^{11}\epsilon_{1,1} - h^{11}_{,1} \dot{\epsilon}_1 + \dot{h}\epsilon_{,1}^1 - h_{,1}\dot{\epsilon}^1.\eqno(84)$$
Consequently $S_2$ is invariant under the transformations of eqs. (81-82) provided surface terms can be neglected when integrating by parts. The right side of eq. (84) does not arise if instead of performing an integration by parts to arrive at $S_2$ in eq. (35), we had performed a gauge transformation on the original action appearing in the two dimensional version of eq. (1).

We can also show that
$$\left\lbrace G(\epsilon, \epsilon_1, \epsilon^1) G(\eta, \eta_1, \eta^1)\right\rbrace \nonumber$$
$$= G\left( 2\left(\epsilon^1\eta_1 - \epsilon_1 \eta^1\right), 
\epsilon\eta_1  - \epsilon_1 \eta_, - \left(\epsilon\eta^1  - \epsilon^1 \eta\right)\right)\eqno(85)$$
so that the generator of the transformation of eqs. (81,82) satisfies a closed algebra. Its structure is the same as that of eqs. (45-47). We thus see that the gauge generator satisfies a local closed algebra with field independent structure constants. In this, the two dimensional Einstein-Hilbert action more closely resembles non-Abelian gauge theory than the four dimensional Einstein-Hilbert action when analyzed using the Dirac-ADM [7-9] analysis of constraints.

We now consider how the two dimensional Einstein-Hilbert action can be treated if $g^{\alpha\beta}$ and $\Gamma_{\mu\nu}^\lambda$ are taken to be the canonical ``position'' variables in place of $h^{\alpha\beta}$ and $\Gamma_{\mu\nu}^\lambda$. This eliminates the need to impose the condition of eq. (10) ``by hand'' in order for the theory to be a metric theory. It also permits supplementing the action $S_2$ with a cosmological term $\Lambda \int d^2x \sqrt{-g}$. This term cannot be expressed in terms of $h^{\alpha\beta}$ in two dimensions.

We now find it convenient to write $S_2$ in the form
$$S_2 = \int d^2x \left[ h^{11} \dot{\xi}_{11} + h^1\dot{\xi}_1 + h\dot{\xi} - \left[\zeta^1\left(h_{,1} + h\xi_1 + 2h^1 \xi_{11}\right)\right.\right.\nonumber$$
$$\left. \left. + \zeta_1 \left(h_{,1}^{11} - 2h^1\xi - h^{11}\xi_1\right) + \zeta\left(h_{,1}^1 - h\xi + h^{11}\xi_{11}\right)\right]\right] ,\eqno(86)$$
where
$$\xi_{11} = \Gamma_{11}^0, \;\;\;\;\;\xi_1 = \Gamma_{01}^0 - \Gamma_{11}^1 ,\;\;\;\;\; \xi = -\Gamma_{01}^1\eqno(87-89)$$
and now $h^{11}$, $h^{1}$ and $h$ are not independent variables but are defined by the equations
$$h^{11} = \sqrt{-g}\, g^{11},\;\;\;\;\;\; h^{1} = \sqrt{-g}\, g^{01},\;\;\;\;\;\;
h = \sqrt{-g}\, g^{00}. \eqno(90-92)$$
In this form, we have the canonical momenta
$$\pi^{11} = \frac{\partial L}{\partial \dot{\xi}_{11}} = \sqrt{-g}\, g^{11},\;\;\;\;\;
\pi^{1} = \frac{\partial L}{\partial \dot{\xi}_{1}} = \sqrt{-g}\, g^{01},\;\;\;\;\;
\pi = \frac{\partial L}{\partial \dot{\xi}} = \sqrt{-g}\, g^{00},\eqno(93-95)$$
$$p_{11} = \frac{\partial L}{\partial \dot{g}^{11}} = 0,\;\;\;\;\;
p_{1} = \frac{\partial L}{\partial \dot{g}^{01}} = 0,\;\;\;\;\;
p = \frac{\partial L}{\partial \dot{g}^{00}} = 0,\eqno(96-98)$$
$$\Pi_1 = \frac{\partial L}{\partial \dot{\zeta}^1} = 0,\;\;\;\;\;
\Pi^{1} = \frac{\partial L}{\partial \dot{\zeta}_{1}} = 0,\;\;\;\;\;
\Pi = \frac{\partial L}{\partial \dot{\zeta}}\;\; .\eqno(99-101)$$
These are all primary constraints. The constraints of eq. (99-101) are all first class. Those of eqs. (93-98) are not all second class, as is apparent from the fact that the matrix formed by the PB of these constraints is of rank four, not six. This is on account of eq. (10) which implies that if (93-95) are satisfied then
$$\Xi_{\pi}^{11} = \pi\pi^{11} - (\pi^1)^2 + 1 = 0.\eqno(102)$$
There are in addition the secondary constraints
$$\tilde{\chi}_1 = \pi_{,1} + \pi\xi_1 + 2\pi^1\xi_{11} = 0,\eqno(103)$$
$$\tilde{\chi}^1 = \pi_{,1}^{11} - 2\pi^1\xi - \pi^{11}\xi_1 = 0,\eqno(104)$$
$$\tilde{\chi} = \pi_{,1}^1 - \pi\xi + \pi^{11}\xi_{11} = 0.\eqno(105)$$
When eq. (102) is satisfied, then the three constraints of eqs. (103-105) are no longer independent as now
$$\pi^{11}\tilde{\chi}_1 + \pi\tilde{\chi}^1 - 2\pi^1\tilde{\chi}  = 0\;.\eqno(106)$$

We could take four of the six constraints of eqs. (93-98) as being second class and the remaining two to be first class; eqs. (99-101) would constitute an additional three first class constraints. Finally, two of the three constraints of eqs. (103-105) would be first class. In total then, there are four second class constraints and seven first class constraints. When these are combined with seven gauge conditions, there are in total $4 + 7 + 7 = 18$ restrictions on the 18 variables $g^{\alpha\beta}$, $\Gamma_{\mu\nu}^\lambda$ and their conjugate momenta.

In place of taking two of the constraints of eq. (93-98) to be first class, we instead find it convenient to take eq. (102) to be first class along with all of the constraints of eqs. (103-105). There are still seven first class constraints and no degrees of freedom. The constraints of eqs. (103-105) satisfy the algebra of eqs. (45-47) and have a vanishing PB with the constraint $\Xi^{11}_\pi$ of eq. (102).

The approach of Castellani [10] can again be used to find the generator of the gauge transformation associated with first class constraints of eqs. (99-105).
Following the approach that led to $G$ in eq. (62), we arrive at
$$\tilde{G} = \int\left[\epsilon\left(\tilde{\chi} - \zeta^1\Pi_1 + \zeta_1\Pi^1\right) + \epsilon_1 \left(\tilde{\chi}^1 - 2\zeta^1\Pi - \zeta\Pi^1\right)\right.\nonumber$$
$$\left. + \epsilon^1 \left(\tilde{\chi}_1 + 2\zeta_1\Pi + \zeta\Pi_1\right) + \dot{\epsilon} \Pi + \dot{\epsilon}_1\Pi^1 + \dot{\epsilon}^1\Pi_1 + \tilde{\epsilon}_{11} \Xi^{11}_\pi\right]dx .\eqno(107)$$
From eq. (107), we recover the transformations of eqs. (63-65), (72-77) with the additional transformations
$$\delta\Gamma_{11}^0 = \tilde{\epsilon}_{11}\pi,\eqno(108)$$
$$\delta\Gamma_{01}^1 = -\tilde{\epsilon}_{11}\pi^{11},\eqno(109)$$
$$\delta\Gamma_{11}^1 = 2\tilde{\epsilon}_{11}\pi^{1},\eqno(110)$$
$$\delta\Gamma_{00}^0 = -\tilde{\epsilon}_{11}\pi^{11}.\eqno(111)$$
With $L$ now given by
$$L = \pi^{11}\dot{\xi}_{11} + \pi^1\dot{\xi}_1 + \pi\dot{\xi} - \left(\zeta^1 \tilde{\chi}_1 + \zeta_1\tilde{\chi}^1 + \zeta\tilde{\chi}\right)\eqno(112)$$
then it follows that
$$\!\!\!\!\!\!\!\!\!\!\left\lbrace L, G\right\rbrace = \epsilon_{,1}^1 \dot{\pi} - \dot{\epsilon}^1 \pi_{,1} + \epsilon_{1,1}\dot{\pi} - \dot{\epsilon}_1\pi^{11}_{\;,1} + \epsilon_{,1}\dot{\pi}^1 - \dot{\epsilon}\pi_{,1}^1\nonumber$$
$$- \frac{d}{dt}\left[\epsilon^1\left(\pi\xi_1 + 2\pi^1\xi_{11}\right)+\epsilon_1\left(-2\pi^1\xi - \pi^{11}\xi_1\right) + \epsilon \left(-\pi\xi + \pi^{11}\xi_{11}\right)\right]
-\tilde{\epsilon}_{11} \frac{d}{dt}(\Xi^{11}_\pi
),\eqno(113)$$
so that provided surface terms can be neglected and eq. (102) is satisfied, the action $S_2$ is invariant.

A more direct approach is to select two of the six constraints following from eqs. (93-98) to be first class and use the remaining four second class constraints to explicitly eliminate four degrees of freedom.  If we take eqs. (94,95, 97, 98) to be second class constraints, we have the strong equations
$$p = 0 = p_1,\eqno(114-115)$$
$$g^{00} = - \frac{\pi^2}{1 - (\pi^1)^2} g^{11},\;\;\;\;\;\;g^{01} = - \frac{\pi\pi^1}{1 - (\pi^1)^2} g^{11}\eqno(116-117)$$
so that $g^{00}$ and $g^{01}$ are expressed in terms of $\pi$, $\pi^1$ and $g^{11}$.

The remaining five primary first class constraints present in eqs. (93-101) can now be written using these strong equations as
$$\Pi_1 = \Pi^1 = \Pi = 0, \;\;\;\;\;p_{11} = 0,\;\;\;\;\; \pi^{11} = \frac{-1+(\pi^1)^2}{\pi}.\eqno(118-122)$$
Furthermore, the action of eq. (86) now can be expressed as
$$S_2 = \int d^2x \left\lbrace\left(\frac{-1+(\pi^1)^2}{\pi}\right)\dot{\xi}_{11} + \pi^1\dot{\xi}_1 + \pi\dot{\xi}\right.\eqno(123)$$
$$-\left[\zeta^1 \left(\pi_{,1} + \pi\xi_1 + 2\pi^1\xi_{11}\right)\right.\nonumber$$
$$+ \zeta_1\left( \left( \frac{-1+(\pi^1)^2}{\pi}\right)_{,1} - 2\pi^1 \xi - \left(
\frac{-1+(\pi^1)^2}{\pi}\right)\xi_1\right)\nonumber$$
$$\left.\left. \left. + \zeta\left( \pi_{,1}^1 - \pi\xi + ( \frac{-1+(\pi^1)^2}{\pi}\right)\xi_{11}\right)\right]\right\rbrace\;\;.\nonumber$$
The first three constraints of eqs. (118-122) are primary constraints that lead to the secondary first class constraints
$$
\tilde \chi_1 = \pi_{,1} + \pi\xi_1 + 2\pi^1\xi_{11},\eqno(124)$$
$$
\tilde \chi^1 = \left( \frac{-1+(\pi^1)^2}{\pi}\right)_{,1} - 2\pi^1\xi - 
\left( \frac{-1+(\pi^1)^2}{\pi}\right)\xi_1,\eqno(125)$$
$$
\tilde \chi = \pi_{,1}^1 - \pi\xi+ \left( \frac{-1+(\pi^1)^2}{\pi}\right)
\xi_{11}.\eqno(126)$$
These constraints are not independent as they satisfy eq. (106) automatically. 
They also obey the algebra of eqs. (45-47). If $\tilde \chi$ and $\tilde 
\chi_1$
are taken to be independent, we then have
$$
\tilde \chi^1 = \frac{1}{\pi}\left( 2\pi^1 \tilde \chi
- \frac{-1+(\pi^1)^2}{\pi} \tilde \chi_1
\right)\eqno(127)$$
as the dependent secondary constraint.

Together, the seven first class constraints (five primary given by eqs. 
(118-122) and two secondary $-\tilde \chi$ and $-\tilde \chi_1$ 
result in a gauge generator
$$G = \int dx\left[ \epsilon^{11}p_{11} + \epsilon_{11} \left(\pi^{11} + \frac{-1+(\pi^1)^2}{\pi}\right) + \epsilon_1 \left(\tilde \chi^1
- \zeta\Pi^1-2\zeta^1\Pi\right)\right.$$
$$ + \epsilon^1\left(\tilde \chi_1
+ 2\zeta_1\Pi + \zeta\Pi_1\right)+\epsilon\left(\tilde \chi
- \zeta^1\Pi_1 + \zeta_1\Pi^1\right) \nonumber$$
$$\left. + \dot{\epsilon}_1 \Pi^1 + \dot{\epsilon}^1\Pi_1 + \dot{\epsilon}\Pi\right] .\eqno(128)$$
In deriving eq. (128) using the approach of ref. [10], we have treated the secondary constraints of eqs. (124-126) as independent though they are related by eq. (127); strictly speaking the approach of ref. [10] assumes that the secondary constraints are not dependent. Despite this, $G$ of eq. (128) is consistent with invariances of $S_2$ in eq. (123).

Keeping in mind eqs. (124-126) we find that $G$ satisfies the algebra of eq. (85). The changes in the dynamical variables induced by $G$ now take the form
$$\delta g^{11} = \epsilon^{11}\eqno(129)$$
(This reflects the fact that $S_2$ in eq. (123) is independent of $g^{11}$.),
$$\delta g^{00} = \epsilon^{11} \frac{g^{00}}{g^{11}} + 2\epsilon g^{00} + 2\epsilon^1 \frac{g^{00}g^{01}}{g^{11}} - 2 \epsilon_1 g^{01},\eqno(130)$$
$$\delta g^{01} = \epsilon^{11} \frac{g^{01}}{g^{11}} + \epsilon g^{01} +  \frac{2\epsilon^1(g^{01})^2}{g^{11}} -  \epsilon_1 g^{11}-\epsilon^1g^{00},\eqno(131)$$
$$\delta \Gamma_{00}^1 = \dot{\epsilon}^1 -\epsilon\Gamma_{00}^1 + \epsilon^1\left( \Gamma_{01}^1 - \Gamma_{00}^0\right),\eqno(132)$$
$$\delta \Gamma_{01}^0 = \dot{\epsilon}_1 + \epsilon\Gamma_{01}^0 - \epsilon_1\left( \Gamma_{01}^1 - \Gamma_{00}^0\right),\eqno(133)$$
$$\delta\Gamma_{01}^1 = \epsilon_{,1}^1 - \epsilon^1\left(\Gamma_{01}^0 - \Gamma_{11}^1\right) - \epsilon\Gamma_{01}^1 - \frac{g^{11}}{g^{00}}\left(\epsilon_{11} - \epsilon\Gamma_{11}^0 - \epsilon_{1,1} - \epsilon_1\left(\Gamma_{01}^0 - \Gamma_{11}^1\right)\right),\eqno(134)$$
$$\delta\Gamma_{11}^0 = \epsilon_{11},\eqno(135)$$
$$\delta\Gamma_{00}^0 = \epsilon_{,1}^1 + \epsilon^1\left(\Gamma_{01}^0 + \Gamma_{11}^1\right) - \epsilon\Gamma_{01}^1 - \frac{g^{11}}{g^{00}}\left(
\epsilon_{11} - \epsilon\Gamma_{11}^0 - \epsilon_{1,1} -\epsilon_1 \left(\Gamma_{01}^0 - \Gamma_{11}^1\right)\right)\nonumber$$
$$- \dot{\epsilon}- 2\epsilon_1\Gamma_{00}^1,\eqno(136)$$
$$\delta\Gamma_{11}^1 = \dot{\epsilon}_1 + \epsilon\Gamma_{01}^0 + \epsilon_1\left(\Gamma_{01}^1 + \Gamma_{00}^0\right) + \epsilon_{,1}-2\epsilon^1\Gamma_{11}^0 + \frac{2g^{01}}{g^{00}}
\left(\epsilon_{11} - \epsilon\Gamma_{11}^0 \right.\nonumber$$
$$ \left. - \epsilon_{1,1} - \epsilon_1 \left(\Gamma_{01}^0 - \Gamma_{11}^1\right)\right).\eqno(137)$$

It is not clear if there is a relationship between the transformations of eqs. (132-137) and those of eqs. (72-77, 108-111), though eqs. (63-65) are consistent with eqs. (129-131).

We now turn to examining how the ADM approach to the canonical analysis of the Einstein-Hilbert action expressed in first order form is related to what has been done above.

\section{The ADM Approach}

The canonical analysis of the four dimensional Einstein-Hilbert action in refs. [7,8] was originally performed using $g_{\alpha\beta}$ as the only dynamical variable; $\Gamma_{\mu\nu}^\lambda$ was given by the Christoffel symbol in this analysis.  (For reviews of this approach, see for example [9,28,29].)

The ADM canonical Hamiltonian is also derived in the first order formalism with $g_{\alpha\beta}$ and $\Gamma_{\mu\nu}^\lambda$ being treated as independent variables in [23,24]. This analysis differs from what appears in section 2 above by using the four secondary constraints of eqs. (22-23) as well as the 26 constraints following from the equations of motion of the variables $\Sigma_i$, $\Sigma_j^i$ and $\Sigma_{jk}^i$ to eliminate in total 30 of the 50 variables appearing in the original Lagrangian (10 components of $g_{\alpha\beta}$ and 40 components of $\Gamma_{\mu\nu}^\lambda$).
However, when these equations are used, the coefficients of the four variables $\Lambda$ and $\Lambda^k$ appearing in eq. (21) all vanish, and hence the Lagrangian analyzed in the ADM approach only contains 50-30-4 = 16 variables; these are $g_{\mu\nu}$ and $\Gamma_{ij}^0$. Four of the components of $g_{\mu\nu}$ appear as Lagrange multipliers in the Hamiltonian; the six remaining components of $g_{\mu\nu}$
have canonical momenta composed of the components of $\Gamma_{ij}^0$. There are thus 20 variables in all in phase space when the momenta $\Pi_A$ conjugate to the Lagrange multipliers $A = (\Lambda , \Lambda^k)$ are considered in addition to the 16 variables appearing explicitly in the Lagrangian. There consequently are four primary constraints (the momenta $\Pi_A$ vanish) and four secondary constraints (the coefficients of the Lagrange multipliers). These eight constraints are all first class; when combined with eight gauge conditions there are 20-8-8=4 independent degrees of freedom in phase space. These are the two propagating modes of the graviton plus their conjugate momenta. The four primary and four secondary first class constraints can be used to form the generator of a gauge transformation [10].

In sections two and three above we have not used equations of motion that are independent of time derivatives to eliminate dynamical variables. Indeed, by using equations of motion associated with first class constraints it is inevitable that some information about the gauge invariance present in the initial Lagrangian would be lost. Furthermore, tertiary constraints will not arise, and if these are first class, even more information is lost. Indeed, having only 16 of the original 50 degrees of freedom left in the Lagrangian after  using the time independent equations of motion means that only by employing the equations of motion in conjunction with the generator of the gauge transformation can the transformation of all 50 fields $g_{\alpha\beta}$ and $\Gamma_{\mu\nu}^\lambda$ be found. (We note that in analyzing the canonical structure of the four dimensional Einstein-Hilbert Lagrangian when expressed in terms of the vierbein and spin connection, it is not necessary to employ equations of motion [12].)

Let us explicitly demonstrate what happens when the approach to the Einstein-Hilbert
Lagrangian in four dimensions used in [23,24] is applied to the action $S_2$ in eq. (35).
In this case, the variables $\zeta_1$, $\zeta^1$ and $\zeta$ do not enter with time derivatives and hence their equations of motion show that $\chi_1$, $\chi^1$ and $\chi$  vanish where these quantities are defined by eqs. (42-44). They also satisfy the restriction that
$$h^{11} \chi_1 + h\chi^1 - 2h^1\chi = (hh^{11} - (h^1)^2)_{,1}.\eqno(138)$$
On account of eq. (138), it is not possible to solve these equations of motion for all of $\pi_{11}$, $\pi_1$ and $\pi$. However, if we use $\chi_1 = \chi^1 = 0$ we find that 
$$\pi_{11} = \frac{1}{2h^1} \left(h_{,1} - h\pi_1\right),\eqno(139)$$
$$\pi = \frac{-1}{2h^1} \left(h_{,1}^{11} + h^{11}\pi_1\right).\eqno(140)$$
Substitution of eqs. (139-140) into eq. (44) leads to
$$\chi = \frac{1}{2h^1} \left(\left(h^1\right)^2 - h h^{11}\right)_{,1}\eqno(141)$$
and 
eq. (35) reduces to
$$S_2 = \int d^2x \left[ \frac{1}{2h^1} \left(h_{,1}\dot{h}^{11} - h^{11}_{,1}\dot{h}\right)
+ \frac{\pi_1}{2h^1} \frac{d}{dt} \left(\left(h^1\right)^2 - hh^{11}\right) - \frac{\zeta}{2h^1}\left(\left(h^1\right)^2 - hh^{11}+k\right)_{,1}\right] .\eqno(142)$$
(We are free to add a constant $k$ in eq. (142) as indicated.)
The first term in eq. (142) cancels against the second term upon integrating by parts. If now we set
$$H_{11} = \frac{\pi_1}{2h^1}\;\;\;\;\; \tilde{H}_{11} = -\left(\frac{\zeta}{2h^1}\right)_{,1}\eqno(143)$$
then eq. (142) reduces to
$$S_2 = \int d^2x \left[ H_{11} \left(2h^1\dot{h}^1 - h^{11}\dot{h} - h\dot{h}^{11}\right)\right. \eqno(144)$$
$$\left. - \tilde{H}_{11} \left(\left(h^1\right)^2 - hh^{11}+k\right) \right].\nonumber$$
We thus have the primary constraints
$$\psi_1 \equiv \pi_1 - 2h^1 H_{11} = 0\, ,\;\;\;\;\;
\psi \equiv \pi + h^{11} H_{11}\, , \;\;\;\;\;
\psi_{11} \equiv \pi_{11} + h H_{11}\, , \eqno(145-147)$$
and
$$\Pi^{11} = 0 = \tilde{\Pi}^{11}\eqno(148-149)$$
for the momenta $\pi_1 \ldots \Pi_{11}$ conjugate to $h^1 \ldots \tilde{H}_{11}$ respectively. There is one secondary constraint
$$\Xi^{11} = (h^1)^2 - hh^{11} + k.\eqno(150)$$
The equations
$$\left\lbrace \psi_{1},\Xi^{11}\right\rbrace = -2h^1,\;\;\;\;\;
\left\lbrace \psi_{1},\Xi^{11}\right\rbrace = h^{11},\;\;\;\;\;
\left\lbrace \psi_{11},\Xi^{11}\right\rbrace = h\eqno(151-153)$$
show that there are four first class constraints which we take to be
$$\Pi^{11} = \tilde{\Pi}^{11} = 0 \eqno(154-155)$$
$$h \psi + h^{11}\psi_{11} + h^1\psi_1 = h\psi - h^{11}\psi_{11} = 0\eqno(156-157)$$
and two second class constraints, for which we choose
$$\psi_1 = \Xi^{11} = 0.\eqno(158-159)$$
Once four gauge conditions are chosen, all ten degrees of freedom associated with $h$, $h^1$, $h^{11}$, $H_{11}$, $\tilde{H}_{11}$ and their conjugate momenta are fixed.

The four first class constraints of eqs. (154-157) are all primary, so the gauge generator is simply [10]
$$G_I = \int \left[ \epsilon_{11}\Pi^{11} + \tilde{\epsilon}_{11}\tilde{\Pi}^{11} + \epsilon_a \left(h\psi + h^{11}\psi_{11} + h^1 \psi_1\right) + \epsilon_b\left(h\psi - h^{11}\psi_{11}\right)\right] dx.\eqno(160)$$
It is also possible to consider the constraints 
$\tilde{\Pi}^{11}$, $\psi_1$, $\psi$ and $\Xi^{11}$ to be first class and $\Pi^{11}$ and $\psi_{11}$ to be second class. This entails using $\psi_{11} = 0$ to express
$$H_{11} = -\frac{-\pi_{11}}{h},\eqno(161)$$
so that we write now
$$\tilde{\psi}_1 = \pi_1 + \frac{2h^1\pi_{11}}{h},\eqno(162)$$
$$\tilde{\psi} = \pi - \frac{h^{11}}{h} \pi_{11} .\eqno(163)$$
Since $\Xi^{11}$ is a secondary constraint associated with the primary constraint $\tilde{\Pi}^{11}$, the gauge generator now becomes
$$G_{II} = \int \left[ \epsilon^1\tilde{\psi}_1 + \epsilon\tilde{\psi} + \tilde{\epsilon}_{11} \Xi^{11} + \dot{\tilde{\epsilon}}_{11}\Pi^{11}\right] dx.\eqno(164)$$
From eqs. (160) and (144), we see that
$$\left\lbrace G_I, L\right\rbrace = \left[ -\epsilon_{11} \frac{d}{dt} - \tilde{\epsilon}_{11} + (2\dot{H}_{11}\epsilon_a)\frac{d}{dt}\right]\Xi^{11},\eqno(165)$$
so that on the constraint surface, $L$ is gauge invariant under the transformation generated by eq. (160). So also, the generator of eq. (164) leaves $S_2$ of eq. (144) invariant on the constraint surface.

\section{Discussion}

In this paper, we have examined the canonical structure of the Einstein-Hilbert action when written in first order form, using the metric and affine connection as independent variables.  The algebraic complexity that arises when applying the Dirac constraint formalism in $d > 2$ dimensions prevents us from fully implementing this way of disentangling the structure of these actions. However, when $d = 2$ it has proved possible to determine all constraints, both using the metric density $h^{\alpha\beta}$ and the metric $g^{\alpha\beta}$ as dynamical variables. In the former case, a new symmetry (given by eqs. (81,82)) is derived. Furthermore, in two dimensions the algebra of constraints is local with field independent structure constants; it is tempting to speculate that these features persist in higher dimensions. The structure of eq. (9) suggests that this formalism may be simpler to implement when using $G_{\mu\nu}^\lambda$ in place of $\Gamma_{\mu\nu}^\lambda$. (The use of $G_{\mu\nu}^\lambda$ will be outlined elsewhere.)

Completing the analysis of the Einstein-Hilbert action in $d > 2$ dimensions along the lines proposed in section 3 is clearly a priority. Another problem that suggests itself is to fully quantize the two dimensional Einstein-Hilbert action when using the first order formalism of section 4 \cite{Gerry}.  Despite the fact that the constraints serve to eliminate all canonical degrees of freedom in this model, the structure of such theories can be of interest [19,20,30]. It would also be worth considering what happens to the canonical structure of the two dimensional Einstein-Hilbert actions if it were supplemented by a cosmological term or matter fields.
Its structures when written in terms of the zweibein and spin connection (as 
in [12]) also merits attention. The use of these variables has been explored 
in [31]; it has been found that tertiary constraints in fact arise in this 
approach, consistent with our own expectations. 

\section{Acknowledgements}

D.G.C.McKeon would like to thank NSERC for financial support and R. and D.
MacKenzie for useful suggestions.

\end{document}